\documentclass[aps,preprint]{revtex4}

\begin{document}
\title{A boundary value problem for the five-dimensional stationary rotating black holes}

\author{Yoshiyuki Morisawa}
\email[]{morisawa@yukawa.kyoto-u.ac.jp}
\affiliation{Yukawa Institute for Theoretical Physics, 
Kyoto University, Kyoto 606-8502, Japan}
\author{Daisuke Ida}
\email[]{d.ida@th.phys.titech.ac.jp}
\affiliation{Department of Physics,
Tokyo Institute of Technology, Tokyo 152-8551, Japan}

\date{January 22, 2004}

\begin{abstract}
We study the boundary value problem for the stationary rotating black hole solutions
to the five-dimensional vacuum Einstein equation.
Assuming the two commuting rotational symmetry and the sphericity of the horizon topology,
we show that the black hole is uniquely characterized by 
the mass, and a pair of the angular momenta.
\end{abstract}

\pacs{}
\preprint{YITP--04--6}

\maketitle

\section{Introduction}
In recent years there has been renewed interest in higher
dimensional black holes in the context of both string theory and brane
world scenario.
In particular, the possibility of black hole production in linear
collider is
suggested~\cite{'tHooft:rb,Banks:1999gd,Giddings:2001bu,Dimopoulos:2001hw}.
Such phenomena play a key role to get insight into 
the structure of space-time;
we might be able to prove the existence of the extra dimensions and
have some information about the quantum gravity.
Since the primary signature of the black hole production in the collider will be 
Hawking emission from the stationary black hole,
the classical equilibrium problem of black holes is an important subject.
The black holes produced in colliders will be small enough compared with the size of the
extra dimensions and generically have angular momenta, 
they will be well approximated by higher dimensional rotating
black hole solutions found by Myers and Perry~\cite{Myers:un}.
The Myers-Perry black hole
which has the event horizon with spherical topology
can be regarded as the higher-dimensional generalization
of the Kerr black hole.
One might expect that such a black hole solution describes
the classical equilibrium state continued from the black hole production event,
if it equips stability and uniqueness like the Kerr black hole in
four-dimensions.
The purpose of this paper is to consider the uniqueness and
nonuniqueness of the rotating black holes in higher dimensions.

The uniqueness theorem states that a four-dimensional black hole with
regular event horizon is characterized only by  mass, angular
momentum and electric charge
~\cite{Carter:hk,Heusler:book}.
Recently, uniqueness and nonuniqueness properties of five or
higher-dimensional black holes are also studied.
Emparan and Reall have found a black ring solution 
of the five-dimensional vacuum Einstein equation, 
which describes a stationary rotating black hole with the event
horizon homeomorphic to $S^2 \times S^1$~\cite{Emparan:2001wn}.
In a certain parameter region, a black ring and
a (Myers-Perry) black hole can carry the
same mass and angular momentum.
This might suggest the nonuniqueness of
higher-dimensional stationary black hole solutions.
For example, Reall~\cite{Reall:2002bh} conjectured the existence of
stationary, asymptotically flat higher-dimensional vacuum black hole
admitting exactly two commuting Killing vector fields although all
known higher dimensional black hole solutions have three or
more Killing vector fields.
In six or higher dimensions, Myers-Perry black hole can have an
arbitrarily large angular momentum for a fixed mass.
The horizon of such black hole highly spreads out in the plane of
rotation and looks like a black brane in the limit where the angular
momentum goes to infinity.
Hence, Emparan and Myers~\cite{Emparan:2003sy} argued that
rapidly rotating black holes are unstable due to the Gregory-Laflamme
instability~\cite{Gregory:vy} and decay to the stationary black holes
with rippled horizons implying
the existence of black holes with less geometric symmetry 
compared with the Myers-Perry black holes.
For supersymmetric black holes and black rings, string theoretical
interpretation are given by Elvang and Emparan~\cite{Elvang:2003mj}.
They showed that the black hole and the black ring with same
asymptotic charges correspond to the different configurations of
branes, giving a partial resolution of the nonuniqueness of
supersymmetric black holes in five dimensions.
On the other hand, we have uniqueness theorems for black holes
at least in the static case~\cite{Hwang:1998,Gibbons:2002bh,Gibbons:2002av,Gibbons:2002ju,Rogatko:2002bt,Rogatko:2003kj}.
Furthermore, the uniqueness of the stationary black holes is supported
by the argument based on linear perturbation of higher dimensional
static black holes~\cite{Kodama:2003jz,Ishibashi:2003ap}.
There exist regular stationary perturbations that fall off at
asymptotic region only for vector perturbation,
and then the number of the independent modes corresponds to the rank of
the rotation group, namely the number of angular momenta carried by
the Myers-Perry black holes~\cite{Kodama:private}.
This suggests that the higher-dimensional stationary black holes have 
uniqueness property in some sense, but some amendments will be required.
Here we consider the possibility of restricted black hole uniqueness which is
consistent with any argument about uniqueness or nonuniqueness.
Though the existence of the black ring solution explicitly violates the
black hole uniqueness, there still be a possibility of black hole uniqueness
for fixed horizon topology~\cite{Kol:2002dr}.
Hence we restrict ourselves to the stationary black holes with
spherical topology.

In this paper, we consider the asymptotically flat, black hole
solution to the five-dimensional vacuum Einstein equation with the
regular event horizon homeomorphic to $S^3$, admitting two commuting
spacelike Killing vector fields and stationary (timelike) Killing
vector field.
The two spacelike Killing vector fields correspond to the
rotations in the ($X^1$-$X^2$)-plane and ($X^3$-$X^4$)-plane in the asymptotic
region ($\{X^{\mu}\}$ are  the asymptotic Cartesian coordinates),
respectively, which are commuting with each other.
Along with the argument by Carter~\cite{Carter:1970},
it is possible to construct a timelike Killing vector field tangent to
the fixed points (namely, axis) of the axi-symmetric Killing vector
field from the given timelike Killing vector field.
Repeating this procedure for each commuting spacelike Killing vector
field, the obtained timelike Killing vector field is also commuting
with both spacelike Killing vector fields.
Hence, it is natural to assume all the three Killing vector fields are
commuting with each other.
The five-dimensional vacuum space-time admitting three commuting
Killing vector fields is described by the nonlinear
$\sigma$-model~\cite{Maison:kx}.
Then the Mazur
identity~\cite{Mazur:1984wz} for this system is derived.
We show that the
five-dimensional black hole solution with regular event horizon of
spherical topology is determined by three parameters under the
appropriate boundary conditions.

The remainder of the paper is organized as follows.
In Section~\ref{sec:metric}, 
we give the field equations for the five-dimensional vacuum
space-time admitting three commuting Killing vector fields.
In Section~\ref{sec:matrix}, we introduce the matrix form of
field equations to clarify the hidden symmetry of this system
following Maison~\cite{Maison:kx}.
Then the Mazur identity which is useful to show the coincidence of two
solutions is derived in Section~\ref{sec:identity}.
In Section~\ref{sec:conditions}, we determine the boundary conditions.
We summarize this paper and make discussions on related matters in
Section~\ref{sec:summary}.

\section{Five-dimensional vacuum space-time admitting three commuting
Killing vector fields}
Assuming the symmetry of space-time, the Einstein equations reduce to
the equations for the scalar fields defined on three-dimensional
space.
Then, we show that the system of the scalar fields is described by a
nonlinear $\sigma$-model.
\subsection{Weyl-Papapetrou metrics}
\label{sec:metric}
We consider the five-dimensional space-time admitting two commuting
Killing vector fields $\xi_I=\partial_I,\,(I=4,5)$.
The metric can be written in the form
\begin{eqnarray}
g &=&
f^{-1}\gamma_{ij} dx^i dx^j
+f_{IJ}(dx^I+w^I_i dx^i)(dx^J+w^J_j dx^j),
\end{eqnarray}
where $i,j=1,2,3$, $f=\det(f_{IJ})$.
The three-dimensional metric $\gamma_{ij}$, the functions $w^I_i$ and
$f_{IJ}$ are independent on the coordinates
$x^I$ ($x^4=\phi$, $x^5=\psi$,
and we will later identify $\xi_4$ and $\xi_5$ as Killing vector
fields corresponding to two independent rotations in the case of
asymptotically flat space-time).
We define the twist potential $\omega_I$ by
\begin{eqnarray}
\omega_{I,\mu} = f\,f_{IJ}\sqrt{|\gamma|}\epsilon_{ij\mu}
 \gamma^{im}\gamma^{jn}\partial_m w^J_n,
\label{eq:def-twist}
\end{eqnarray}
where $\mu=1,\cdots,5$, $\gamma=\det(\gamma_{ij})$,
$\gamma^{ij}$ is the inverse metric of $\gamma_{ij}$, and
$\epsilon_{\lambda\mu\nu}$ denotes the totally skew-symmetric symbol
such that $\epsilon_{123}=1,\epsilon_{I\mu\nu}=0$.
Then the vacuum Einstein equation reduces to the field equations for
the five scalar fields $f_{IJ}$ and $\omega_{I}$ defined on the
three-dimensional space:
\begin{eqnarray}
D^2 f_{IJ} &=&
f^{KL}Df_{IK}\cdot Df_{JL} -f^{-1}D\omega_I \cdot D\omega_J,
\label{eq:EOM-f}
\\
D^2 \omega_I &=&
f^{-1}Df\cdot D\omega_I +f^{JK}Df_{IJ}\cdot D\omega_K,
\label{eq:EOM-omega}
\end{eqnarray}
and the Einstein equations on the three-dimensional space:
\begin{eqnarray}
{}^{(\gamma)}R_{ij}
&=&
{1 \over 4}f^{-2}f_{,i}f_{,j}
+{1 \over 4}f^{IJ}f^{KL}f_{IK,i}f_{JL,j}
+{1 \over 2}f^{-1}f^{IJ}\omega_{Ii}\omega_{Jj},
\end{eqnarray}
where $D$ is the covariant derivative with respect to the three-metric
$\gamma_{ij}$ and the dot denotes the inner product determined by
$\gamma_{ij}$.

Here we assume the existence of another Killing vector field
$\xi_3=\partial_3$ which commutes with the other Killing vectors
as $[\xi_3,\xi_I]=0$
(we will later identify the $\xi_3$ as the stationary Killing vector
field in the case of asymptotically flat space-time).
Then the metric can be written in the Weyl-Papapetrou--type
form~\cite{Ida:2003wv}
\begin{eqnarray}
g &=&
f^{-1}e^{2\sigma}(d\rho^2+dz^2)-f^{-1}\rho^2 dt^2
+f_{IJ}(dx^I+w^I dt)(dx^J+w^J dt),
\label{eq:weyl-metric}
\end{eqnarray}
where we denote $x^3=t$,
and all the metric functions depend only on $\rho$ and $z$.
Once the five scalar fields $f_{IJ}, \omega_I$ are determined, the
other metric functions $\sigma$ and $w^I$ are obtained by solving the
following partial derivative equations:
\begin{eqnarray}
{2 \over \rho}\sigma_{,\rho} &=&
{1 \over 4} f^{-2}[(f_{,\rho})^2-(f_{,z})^2]
+{1 \over 4} f^{IJ}f^{MN}(f_{IM,\rho}f_{JN,\rho}-f_{IM,z}f_{JN,z})
\nonumber\\&&
+{1 \over 2} f^{-1}f^{IJ}
(\omega_{I,\rho}\omega_{J,\rho}-\omega_{I,z}\omega_{J,z}),
\\
{1 \over \rho}\sigma_{,z} &=&
{1 \over 4} f^{-2} f_{,\rho} f_{,z}
+{1 \over 4} f^{IJ}f^{MN} f_{IM,\rho} f_{JN,z}
+{1 \over 2} f^{-1}f^{IJ} \omega_{I,\rho} \omega_{J,z},
\\
w^I_{,\rho} &=&
\rho f^{-1} f^{IJ} \omega_{J,z},
\\
w^I_{,z} &=&
-\rho f^{-1} f^{IJ} \omega_{J,\rho}.
\end{eqnarray}
The $f_{IJ}$ and $\omega_I$ are given by axi-symmetric solution
of the field equations~(\ref{eq:EOM-f}) and~(\ref{eq:EOM-omega}) on
the abstract flat three-space with the metric
\begin{eqnarray}
\gamma &=& d\rho^2+dz^2+\rho^2 d\varphi^2.
\label{eq:flat-metric}
\end{eqnarray}
Thus the system is described by the action
\begin{eqnarray}
S&=&\int d\rho dz \,\rho
\left[{1 \over 4}f^{-2}(\partial f)^2
+{1 \over 4}f^{IJ}f^{KL}\partial f_{IK} \cdot \partial f_{JL}
+{1 \over 2}f^{-1}f^{IJ}\partial\omega_I \cdot \partial\omega_J \right].
\label{eq:eff-action}
\end{eqnarray}

\subsection{Matrix representation}
\label{sec:matrix}
The action~(\ref{eq:eff-action}) is invariant under the global
$SL(3,{\bf R})$ transformations as shown by Maison~\cite{Maison:kx}.
Instead of the nonlinear representation by the scalar fields $f_{IJ}$
and $\omega_I$, we introduce the $SL(3,{\bf R})$ matrix field $\Phi$ as
\begin{eqnarray}
\Phi &=&
\left(
\begin{array}{rrr}
f^{-1} & -f^{-1}\omega_{\phi} & -f^{-1}\omega_{\psi}
\\
-f^{-1}\omega_{\phi} & f_{\phi\phi}+f^{-1}\omega_{\phi}\omega_{\phi} & f_{\phi\psi}+f^{-1}\omega_{\phi}\omega_{\psi}
\\
-f^{-1}\omega_{\psi} & f_{\phi\psi}+f^{-1}\omega_{\phi}\omega_{\psi} & f_{\psi\psi}+f^{-1}\omega_{\psi}\omega_{\psi}
\\
\end{array}
\right),
\label{eq:def-Phi}
\end{eqnarray}
which is symmetric (${}^{t}\Phi=\Phi$) and unimodular ($\det\Phi=1$).
$\Phi$ transforms as a covariant, symmetric, second-rank tensor fields
under global $SL(3,{\bf R})$ transformations.
When the Killing vector fields $\xi_{\phi}$ and $\xi_{\psi}$ are
spacelike, all the eigenvalues of $\Phi$ are real and positive.
Therefore, there is an $SL(3,{\bf R})$ matrix field $g$ which is a
square root of the matrix field $\Phi$, namely
\begin{eqnarray}
\Phi = g \, {}^{t}g.
\label{eq:squareroot}
\end{eqnarray}
This square root matrix $g$ is determined upto global $SO(3)$ rotation
because the rotation $g \mapsto g\Lambda$ for any $\Lambda \in SO(3)$
is canceled by $\Lambda^{-1}={}^{t}\Lambda$.
Since any $SL(3,{\bf R})$ matrix field $g$ conversely defines a
symmetric and unimodular matrix field by $\Phi=g \, {}^{t}g$, the
matrix $\Phi$ defines a map from two-dimensional $\rho$-$z$-half plane
(base space) to the coset space $SL(3,{\bf R})/SO(3)$.

The inverse matrix of $\Phi$ is explicitly given by
\begin{eqnarray}
\Phi^{-1} &=&
\left(
\begin{array}{ccc}
f+f^{IJ}\omega_I \omega_J & f^{\phi J}\omega_J & f^{\psi J}\omega_J
\\
f^{\phi J}\omega_J & f^{\phi\phi} & f^{\phi\psi}
\\
f^{\psi J}\omega_J & f^{\phi\psi} & f^{\psi\psi}
\end{array}
\right),
\end{eqnarray}
and transforms as a second rank contravariant tensor field on the base
space.

The current matrix defined by
\begin{eqnarray}
J_i = \Phi^{-1}\partial_i \Phi
\end{eqnarray}
linearly transforms according to the adjoint representation of
$SL(3,{\bf R})$.
This current is conserved, namely every element of $D_{i}J^{i}$
independently vanishes due to the field equations~(\ref{eq:EOM-f})
and~(\ref{eq:EOM-omega}).

The action~(\ref{eq:eff-action}) can be expressed in terms of $J_i$ or
$\Phi$ as
\begin{eqnarray}
S&=& {1 \over 4}\int d\rho dz \,\rho{\rm tr}(J_i J^i),
\\
&=& {1 \over 4}\int d\rho dz \,\rho
{\rm tr}(\Phi^{-1}\partial_i \Phi \Phi^{-1}\partial^i \Phi).
\end{eqnarray}
This action takes a nonlinear $\sigma$-model form.

\section{Mazur identity}
\label{sec:identity}
Let us consider two different sets of the field configurations
$\Phi_{[0]}$ and $\Phi_{[1]}$ satisfying the field
equations~(\ref{eq:EOM-f}) and~(\ref{eq:EOM-omega}).
To show the coincidence of the two solutions,
we will derive the Mazur identity for the nonlinear
$\sigma$-model on the symmetric space $SL(3,{\bf R})/SO(3)$

A bull's eye ${}^{\odot}$ denotes the difference between the value of
functional obtained from the field configuration $\Phi_{[1]}$ and
value obtained from $\Phi_{[0]}$, {\it e.g.},
\begin{eqnarray}
\stackrel{\odot}{J}\!{}^i = {J}^i_{[1]}-{J}^i_{[0]}
=\Phi^{-1}_{[1]}\partial^{i}\Phi_{[1]}-\Phi^{-1}_{[0]}\partial^{i}\Phi_{[0]}.
\end{eqnarray}
The deviation matrix $\Psi$ is defined by
\begin{eqnarray}
\Psi = \stackrel{\odot}{\Phi}\Phi^{-1}_{[0]}
=\Phi_{[1]}\Phi^{-1}_{[0]}-{\bf 1},
\end{eqnarray}
where ${\bf 1}$ is the unit matrix.
The deviation $\Psi$ vanishes if and only if the two sets
of field configurations ($[1]$ and $[0]$) coincide with each other.
Differentiating $\Psi$,
\begin{eqnarray}
D^{i}\Psi = \Phi_{[1]} \stackrel{\odot}{J}\!{}^{i}\, \Phi^{-1}_{[0]},
\end{eqnarray}
and taking divergence, we obtain
\begin{eqnarray}
D_{i}(D^{i}\Psi) =
\Phi_{[1]} D_{i}\stackrel{\odot}{J}\!{}^{i}\, \Phi^{-1}_{[0]}
+\Phi_{[1]} \left\{
J_{[1]i}J_{[1]}^{i}-2J_{[1]i}J_{[0]}^{i}+J_{[0]i}J_{[0]}^{i}
\right\}\Phi^{-1}_{[0]}.
\label{eq:Psiii}
\end{eqnarray}
Due to the current conservation $D_{i}J^{i}=0$, the first term of the
right hand side of Eq.~(\ref{eq:Psiii}) vanishes.
Since ${}^{t}\!J^i = \Phi J^i \Phi^{-1}$, the second term of the right
hand side can be rewritten as
\begin{eqnarray}
\Phi_{[1]} \left\{
J_{[1]i}J_{[1]}^{i}-2J_{[1]i}J_{[0]}^{i}+J_{[0]i}J_{[0]}^{i}
\right\}\Phi^{-1}_{[0]}
&=&
\Phi_{[1]} \left(
J_{[1]}^{i} \stackrel{\odot}{J}\!{}_{i}\,
-\stackrel{\odot}{J}\!{}_{i}\,J_{[0]}^{i}
\right)\Phi^{-1}_{[0]}
\\
&=&
{}^{t}\!J_{[1]}^{i} \Phi_{[1]} \stackrel{\odot}{J}\!{}_{i}\, \Phi^{-1}_{[0]}
-\Phi_{[1]} \stackrel{\odot}{J}\!{}_{i}\,\Phi^{-1}_{[0]}\,{}^{t}\!J_{[0]}^{i}.
\end{eqnarray}
Then taking trace, we obtain the identity
\begin{eqnarray}
(D_{i}D^{i}{\rm tr}\Psi) =
{\rm tr}\left\{
{}^{t}\!\stackrel{\odot}{J}\!{}^{i}\, \Phi_{[1]}
\stackrel{\odot}{J}\!{}_{i}\, \Phi^{-1}_{[0]}
\right\}.
\label{eq:Mazure1}
\end{eqnarray}
Since $D$ is covariant derivative with respect to the abstract flat
three-metric~(\ref{eq:flat-metric}) and all quantities are independent 
on $\varphi$, the above identity~(\ref{eq:Mazure1}) is
\begin{eqnarray}
\partial_{a}(\rho \partial^{a}{\rm tr}\Psi) =
\rho h_{ab}{\rm tr} \left\{
{}^{t}\!\stackrel{\odot}{J}\!{}^{a}\, \Phi_{[1]}
\stackrel{\odot}{J}\!{}^{b}\, \Phi^{-1}_{[0]}
\right\},
\label{eq:Mazur2}
\end{eqnarray}
where $h_{ab}$ is the flat two-dimensional metric
\begin{eqnarray}
h = d\rho^2 + dz^2.
\end{eqnarray}
Integrating Eq.~(\ref{eq:Mazur2}) over the relevant  region
$\Sigma=\{(\rho,z)|\rho \ge 0\}$ in $\rho$-$z$ plane, and using
Green's theorem, we find
\begin{eqnarray}
\oint_{\partial\Sigma} \rho \partial^{a} {\rm tr}\Psi dS_{a}
= \int_{\Sigma} \rho h_{ab}{\rm tr} \left\{
{}^{t}\!\stackrel{\odot}{J}\!{}^{a}\, \Phi_{[1]}
\stackrel{\odot}{J}\!{}^{b}\, \Phi^{-1}_{[0]}
\right\}d\rho dz,
\label{eq:Mazur3}
\end{eqnarray}
where the boundary $\partial\Sigma$ is corresponding to the horizon,
the two planes of rotation and infinity.
Since the matrix $\Phi$ has the square root matrix $g$ as
Eq.~(\ref{eq:squareroot}), the integrand of the right hand side of
Eq.~(\ref{eq:Mazur3}) is written by
\begin{eqnarray}
\rho h_{ab}{\rm tr} \left\{
{}^{t}\!\stackrel{\odot}{J}\!{}^{a}\, \Phi_{[1]}
\stackrel{\odot}{J}\!{}^{b}\, \Phi^{-1}_{[0]}
\right\}
=
\rho h_{ab}{\rm tr} \left\{
g^{-1}_{[0]}\, {}^{t}\! \stackrel{\odot}{J}\!{}^{a}\, g_{[1]}
\,{}^{t}\! g_{[1]} \stackrel{\odot}{J}\!{}^{b}\, {}^{t}\!g^{-1}_{[0]}
\right\}
\end{eqnarray}
Thus, we obtain the Mazur identity
\begin{eqnarray}
\oint_{\partial\Sigma} \rho \partial^{a} {\rm tr} \Psi dS_{a}
= \int_{\Sigma} \rho h_{ab}
{\rm tr} \left\{ {\cal M}^{a}\, {}^{t}\!{\cal M}^{b} \right\}d\rho dz,
\label{eq:Mazur4}
\end{eqnarray}
where the matrix ${\cal M}$ is defined by
\begin{eqnarray}
{\cal M}^{a} =
g^{-1}_{[0]}\, {}^{t}\!\stackrel{\odot}{J}\!{}^{a}\, g_{[1]}.
\end{eqnarray}
When the current difference $\stackrel{\odot}{J}\!{}^{a}$ is
not zero, the right hand side of the identity~(\ref{eq:Mazur4}) is
positive.
Hence we must have $\stackrel{\odot}{J}\!{}^{a}=0$ if the boundary
conditions under which the left hand side of Eq.~(\ref{eq:Mazur4})
vanishes are imposed at $\partial\Sigma$.
Then the difference $\Psi$ is a constant matrix over the region
$\Sigma$.
The limiting value of $\Psi$ is zero on at least one part of the
boundary $\partial\Sigma$ is sufficient to obtain the coincidence of
two solutions $\Phi_{[0]}$ and $\Phi_{[1]}$.

\section{Boundary conditions and coincidence of solutions}
\label{sec:conditions}
When one use the Mazur identity, the boundary conditions for the
fields $\Phi$ ({\it i.e.}, $f_{IJ}$ and $\omega_I$) are needed at the
infinity, the two planes of rotation and the horizon.
We will require asymptotic flatness, regularity at the two planes of
rotation, and regularity at the spherical horizon.
Under these conditions, the Mazur identity shows that the
coincidence of the solutions.

An asymptotically flat space-time with mass $M={3\pi m / 8G}$,
angular momenta $J_{\phi}={\pi ma / 4G}$
and $J_{\psi}={\pi mb / 4G}$ (where we restrict ourselves to
the case in which $m>a^2+b^2+2|ab|$) has metric as the following form:
\begin{eqnarray}
g &=& -\left[1-{m \over r^2}+O(r^{-3})\right] dt^2
-\left[{2ma \over r^{4}}+O(r^{-5})\right] dt(ydx-xdy)
\nonumber\\&&
-\left[{2mb \over r^{4}}+O(r^{-5})\right] dt(wdz-zdw)
\nonumber\\
&&
+\left[1+{m \over 2 r^2}+O(r^{-3})\right] [dx^2+dy^2+dz^2+dw^2].
\end{eqnarray}
Here introducing the coordinates
\begin{eqnarray}
x &=& \sqrt{r^2+a^2} \sin\theta \cos[\bar{\phi}-\tan^{-1}(a/r)],\\
y &=& \sqrt{r^2+a^2} \sin\theta \sin[\bar{\phi}-\tan^{-1}(a/r)],\\
z &=& \sqrt{r^2+b^2} \cos\theta \cos[\bar{\psi}-\tan^{-1}(b/r)],\\
w &=& \sqrt{r^2+b^2} \cos\theta \sin[\bar{\psi}-\tan^{-1}(b/r)],
\end{eqnarray}
and proceeding further coordinate transformations
\begin{eqnarray}
d\bar{\phi} &=& d\phi-{a \over r^2+a^2} dr,\\
d\bar{\psi} &=& d\psi-{b \over r^2+b^2} dr,
\end{eqnarray}
then one obtains
\begin{eqnarray}
g &=& -\left[1-{m \over r^2}+O(r^{-3})\right] dt^2
+\left[{2ma(r^2+a^2) \over r^{4}}\sin^2\theta+O(r^{-3})\right] dt d\phi
\nonumber\\&&
+\left[{2mb(r^2+b^2) \over r^{4}}\cos^2\theta+O(r^{-3})\right] dt d\psi
\nonumber\\
&&
+\left[1+{m \over 2 r^2}+O(r^{-3})\right]\times
\biggl[
{r^2+a^2\cos^2\theta+b^2\sin^2\theta \over (r^2+a^2)(r^2+b^2)} r^2 dr^2
\nonumber\\
&& \quad
+(r^2+a^2\cos^2\theta+b^2\sin^2\theta) d\theta^2
+(r^2+a^2)\sin^2\theta d\phi^2+(r^2+b^2)\cos^2\theta d\psi^2
\biggr].
\label{eq:asympt-metric}
\end{eqnarray}
Here the metric~(\ref{eq:asympt-metric}) admits two orthogonal
planes of rotation $\theta=\pi/2$ and $\theta=0$, which are
specified by the azimuthal angles $\phi$ and $\psi$, respectively.
The planes $\theta=0$ and $\theta=\pi/2$ are invariant under the
rotation with respect to the Killing vector fields $\partial_\phi$ and
$\partial_\psi$, respectively.
Both angles $\phi$ and $\psi$ have period $2\pi$.
Comparing the asymptotic form~(\ref{eq:asympt-metric})
with the Weyl-Papapetrou form~(\ref{eq:weyl-metric}),
we derive boundary conditions.

The regularity on invariant planes requires
\begin{eqnarray}
g_{\phi\phi} = f_{\phi\phi} &=& \sin^2\theta \tilde{f}_{\phi\phi},\\
g_{\psi\psi} = f_{\psi\psi} &=& \cos^2\theta \tilde{f}_{\psi\psi},\\
g_{\phi\psi} = f_{\phi\psi} &=& \sin^2\theta \cos^2\theta \tilde{f}_{\phi\psi},
\end{eqnarray}
where the quantities with tilde are regular at both the invariant
planes and the black hole horizon.

The asymptotic behavior of $\tilde{f}_{\phi\phi}$ and
$\tilde{f}_{\psi\psi}$ are derived from Eq.~(\ref{eq:asympt-metric}),
and $\tilde{f}_{\phi\psi}$ is at most $O(r^{-1})$ since Killing vectors
$\partial_{\phi}$ and $\partial_{\psi}$ are asymptotically orthogonal.
\begin{eqnarray}
\tilde{f}_{\phi\phi} &=& r^2+a^2+{m \over 2}+O(r^{-1}),\\
\tilde{f}_{\psi\psi} &=& r^2+b^2+{m \over 2}+O(r^{-1}),\\
\tilde{f}_{\phi\psi} &=& O(r^{-1}).
\end{eqnarray}

Since $f_{\phi\psi}$ is negligible as compared with $f_{\phi\phi}$ and 
$f_{\psi\psi}$ in the asymptotic region,
the leading terms of $g_{t\phi}$ and $g_{t\psi}$ are
$f_{\phi\phi}w^{\phi}$ and $f_{\psi\psi}w^{\psi}$, respectively.
Then, we have
\begin{eqnarray}
f_{\phi\phi}w^{\phi}
&=& {ma \sin^2\theta \over r^2} + O(r^{-3}),
\\
f_{\psi\psi}w^{\psi}
&=& {mb \cos^2\theta \over r^2} + O(r^{-3}).
\end{eqnarray}
Thus we obtain
\begin{eqnarray}
w^{\phi} &=& {ma \over r^4} + O(r^{-5}),
\\
w^{\psi} &=& {mb \over r^4} + O(r^{-5}).
\end{eqnarray}

Similarly, we have
\begin{eqnarray}
g_{tt}
&=& -f^{-1}\rho^2
+f_{\phi\phi}w^{\phi}w^{\phi} +2f_{\phi\psi}w^{\phi}w^{\psi}
+f_{\psi\psi}w^{\psi}w^{\psi}
\\
&=&
-1 + {m \over r^2} + O(r^{-3}).
\end{eqnarray}
Here $O(r^{-2})$ term must come from $-f^{-1}\rho^2$ term since the
$w^{I}$ are $O(r^{-4})$.
Therefore $\rho$ behaves as
\begin{eqnarray}
\rho^2 = \left[r^4+(a^2+b^2)r^2+O(r)\right] \sin^2\theta \cos^2\theta.
\end{eqnarray}

$\rho^2$ does not only vanish at $\phi$-invariant plane ($\sin\theta=0$)
and $\psi$-invariant plane ($\cos\theta=0$), but also vanishes at the
horizon due to the form of the metric~(\ref{eq:weyl-metric}).
Since the horizon has topology of $S^3$, let us introduce
the spheroidal coordinates on $\Sigma$ as
\begin{eqnarray}
z &=& \lambda\mu,\\
\rho^2 &=& (\lambda^2-c^2)(1-\mu^2),
\label{eq:def-lambda}
\end{eqnarray}
where $\mu = \cos2\theta$.
Then the relevant region is
$\Sigma=\{(\lambda,\mu)|\lambda \ge c, -1\le \mu \le 1\}$.
The boundaries $\lambda=c$, $\lambda=+\infty$, $\mu=1$ and $\mu=-1$
correspond to the horizon, the infinity, the $\phi$-invariant plane
and the $\psi$-invariant plane, respectively.
In these coordinates, the two-dimensional metric on $\Sigma$ is given
by
\begin{eqnarray}
h = d\rho^2+dz^2 = (\lambda^2-c^2\mu^2)
\left({d\lambda^2 \over \lambda^2-c^2}+{d\mu^2 \over 1-\mu^2}\right).
\end{eqnarray}
The boundary integral in the left hand side of the Mazur
identity~(\ref{eq:Mazur4}) is explicitly written as
\begin{eqnarray}
\oint_{\partial\Sigma} \rho \partial^{a}{\rm tr}\Psi dS_{a}
&=&
\int_{c}^{\infty} d\lambda \left.\left(
\sqrt{h_{\lambda\lambda} \over h_{\mu\mu}} \rho
{\partial{\rm tr}\Psi \over \partial\mu}
\right)\right|_{\mu=-1}
+
\int_{-1}^{+1} d\mu \left.\left(
\sqrt{h_{\mu\mu} \over h_{\lambda\lambda}} \rho
{\partial{\rm tr}\Psi \over \partial\lambda}
\right)\right|_{\lambda=\infty}
\nonumber\\&&
+
\int_{\infty}^{c} d\lambda \left.\left(
\sqrt{h_{\lambda\lambda} \over h_{\mu\mu}} \rho
{\partial{\rm tr}\Psi \over \partial\mu}
\right)\right|_{\mu=+1}
+
\int_{+1}^{-1} d\mu \left.\left(
\sqrt{h_{\mu\mu} \over h_{\lambda\lambda}} \rho
{\partial{\rm tr}\Psi \over \partial\lambda}
\right)\right|_{\lambda=c},
\label{eq:boundary_integral}
\end{eqnarray}
where
\begin{eqnarray}
{\partial{\rm tr}\Psi \over \partial x^{a}} =
{\partial \over \partial x^{a}}\left[
f^{-1}_{[1]}\left(-\stackrel{\odot}{f}
+f^{IJ}_{[0]}\stackrel{\odot}{\omega}{}_{I}\stackrel{\odot}{\omega}{}_{J}
\right)
+f^{IJ}_{[0]}\stackrel{\odot}{f}{}_{IJ}
\right],
\quad \mbox{for $x^a=\lambda,\mu$.}
\end{eqnarray}
Here the relation between $\lambda$ and $r$ is given by
\begin{eqnarray}
\lambda = {r^2 \over 2}+{a^2+b^2 \over 4}+O(r^{-1}),
\end{eqnarray}
or
\begin{eqnarray}
r = \sqrt{2}\lambda^{1/2}
\left[1-{a^2+b^2 \over 8\lambda}+O(\lambda^{-3/2})\right].
\end{eqnarray}

The boundary conditions for $f_{IJ}$ are summarized as follows:
\begin{center}
\begin{tabular}{cccccc}
\hline\hline
&
$\phi$-invariant plane &
$\psi$-invariant plane &
horizon &
infinity
\\
&
$\mu \to +1$ &
$\mu \to -1$ &
$\lambda \to c$ &
$\lambda \to +\infty$
\\
\hline
$\tilde{f}_{\phi\phi}$
 & $O(1)$ & $O(1)$ & $O(1)$ & $2\lambda+(a^2-b^2+m)/2+O(\lambda^{-1/2})$
\\
$\tilde{f}_{\phi\psi}$
 & $O(1)$ & $O(1)$ & $O(1)$ & $O(\lambda^{-1/2})$
\\
$\tilde{f}_{\psi\psi}$
 & $O(1)$ & $O(1)$ & $O(1)$ & $2\lambda+(b^2-a^2+m)/2+O(\lambda^{-1/2})$
\\
\hline\hline
\end{tabular}
\end{center}
where
\begin{eqnarray}
f_{\phi\phi} &=& {(1-\mu) \over 2}\tilde{f}_{\phi\phi},\\
f_{\phi\psi} &=& {(1-\mu)(1+\mu) \over 4}\tilde{f}_{\phi\psi},\\
f_{\psi\psi} &=& {(1+\mu) \over 2}\tilde{f}_{\psi\psi}.
\end{eqnarray}

Next, let us derive the boundary conditions for the twist potentials.
By the definition of twist potentials, Eq.~(\ref{eq:def-twist}),
\begin{eqnarray}
{\partial\omega_{\phi} \over \partial\lambda}
= -{f\,f_{\phi J} \over \lambda^2-c^2}{\partial w^{J} \over \partial\mu},
&\quad&
{\partial\omega_{\phi} \over \partial\mu}
= {f\,f_{\phi J} \over 1-\mu^2}{\partial w^{J} \over \partial\lambda},
\\
{\partial\omega_{\psi} \over \partial\lambda}
= -{f\,f_{\psi J} \over \lambda^2-c^2}{\partial w^{J} \over \partial\mu},
&\quad&
{\partial\omega_{\psi} \over \partial\mu}
= {f\,f_{\psi J} \over 1-\mu^2}{\partial w^{J} \over \partial\lambda}.
\end{eqnarray}
From the $\mu$ dependence of $f_{IJ}$, the $\mu$ dependence of the
derivatives of the twist potentials are given as follows:
\begin{eqnarray}
{\partial\omega_{\phi} \over \partial\lambda}=
{\partial\omega_{\phi} \over \partial\mu}=
{\partial\omega_{\psi} \over \partial\lambda}=0
\quad\mbox{at $\mu=+1$},&&
{\partial\omega_{\psi} \over \partial\mu}
\mbox{ does not have } (1-\mu) \mbox{ as a factor},
\\
{\partial\omega_{\psi} \over \partial\lambda}=
{\partial\omega_{\psi} \over \partial\mu}=
{\partial\omega_{\phi} \over \partial\lambda}=0
\quad\mbox{at $\mu=-1$},&&
{\partial\omega_{\phi} \over \partial\mu}
\mbox{ does not have } (1+\mu) \mbox{ as a factor}.
\end{eqnarray}

In the asymptotic region ($\lambda \to +\infty$), the derivatives of
the twist potentials behave as
\begin{eqnarray}
{\partial\omega_{\phi} \over \partial\lambda}
&=& O(\lambda^{-3/2}),
\\
{\partial\omega_{\phi} \over \partial\mu}
&=& -{ma \over 2}(1-\mu) +O(\lambda^{-1/2}).
\end{eqnarray}
Thus we obtain
\begin{eqnarray}
\omega_{\phi} = -{ma\over4}\mu(2-\mu) + (1-\mu)^2(1+\mu) O(\lambda^{-1/2}),
\end{eqnarray}
and similarly
\begin{eqnarray}
\omega_{\psi} = -{mb\over4}\mu(2+\mu) + (1-\mu)(1+\mu)^2 O(\lambda^{-1/2}).
\end{eqnarray}

Then, of course, the condition that $\omega_{I}$ are regular on the
horizon is required.

The boundary conditions for $\omega_{I}$ are summarized as follows:
\begin{center}
\begin{tabular}{cccccc}
\hline\hline
&
$\phi$-invariant plane &
$\psi$-invariant plane &
horizon &
infinity
\\
&
$\mu \to +1$ &
$\mu \to -1$ &
$\lambda \to c$ &
$\lambda \to +\infty$
\\
\hline
$\tilde{\omega}_{\phi}$
 & $O\left((1-\mu)^2\right)$ & $O(1+\mu)$ & $O(1)$ & $O(\lambda^{-1/2})$
\\
$\tilde{\omega}_{\psi}$
 & $O(1-\mu)$ & $O\left((1+\mu)^2\right)$ & $O(1)$ & $O(\lambda^{-1/2})$
\\
\hline\hline
\end{tabular}
\end{center}
where
\begin{eqnarray}
\omega_{\phi} &=& -{ma \over 4}\mu(2-\mu)+\tilde{\omega}_{\phi},
\\
\omega_{\psi} &=& -{mb \over 4}\mu(2+\mu)+\tilde{\omega}_{\psi}.
\end{eqnarray}

The behavior of the following quantities which appear in the boundary
integral~(\ref{eq:boundary_integral}) are easily calculated as follows.
\begin{center}
\begin{tabular}{cccccc}
\hline\hline
&
$\phi$-invariant plane &
$\psi$-invariant plane &
horizon &
infinity
\\
&
$\mu \to +1$ &
$\mu \to -1$ &
$\lambda \to c$ &
$\lambda \to +\infty$
\\
\hline
$\partial{\rm tr}\Psi / \partial\lambda$
 & --- & --- & O(1) & $O(\lambda^{-5/2})$
\\
$\partial{\rm tr}\Psi / \partial\mu$
 & $O(1)$ & $O(1)$ & --- & ---
\\
$\rho$
 & $O(\sqrt{1-\mu})$ & $O(\sqrt{1+\mu})$
 & $O(\sqrt{\lambda-c})$ & $O(\lambda)$
\\
$\sqrt{h_{\mu\mu} / h_{\lambda\lambda}}$
 & --- & --- & $O(\sqrt{\lambda-c})$ & $O(\lambda)$
\\
$\sqrt{h_{\lambda\lambda} / h_{\mu\mu}}$
 & $O(\sqrt{1-\mu})$ & $O(\sqrt{1+\mu})$ & --- & ---
\\
\hline\hline
\end{tabular}
\end{center}
Then, the boundary integral~(\ref{eq:boundary_integral}) vanishes.
The difference matrix $\Psi$ is constant and has asymptotic behavior
as
\begin{eqnarray}
\Psi \to \left(
\begin{array}{ccc}
O(\lambda^{-3/2}) & O(\lambda^{-7/2}) & O(\lambda^{-7/2})\\
O(\lambda^{-1/2}) & O(\lambda^{-3/2}) & O(\lambda^{-3/2})\\
O(\lambda^{-1/2}) & O(\lambda^{-3/2}) & O(\lambda^{-3/2})\\
\end{array}
\right),
\quad (\lambda\to +\infty).
\end{eqnarray}
$\Psi$ vanishes at the infinity, and then $\Psi$ is zero over
$\Sigma$.
Thus, the two configurations $\Phi_{[0]}$ and $\Phi_{[1]}$ coincide
with each other.

\section{Summary and Discussion}
\label{sec:summary}
We show uniqueness of the asymptotically flat, black hole solution to
the five-dimensional vacuum Einstein equation with the regular event
horizon homeomorphic to $S^3$, admitting two commuting spacelike
Killing vector fields and stationary Killing vector field.
The solution of this system is determined by only three asymptotic
charges, the mass $M={3\pi m / 8G}$ and the two angular momenta
$J_{\phi}={\pi m a / 4G}$ and $J_{\psi}={\pi m b / 4G}$.
The five-dimensional Myers-Perry black hole solution is unique in this
class.

The vacuum black ring solution fulfills above conditions other than
that on the topology of the horizon.
There exist two black ring solutions which have same mass and angular
momentum, which means uniqueness property fails for the
$S^2 \times S^1$ event horizon.
It is intriguing to investigate how this nonuniqueness occurs.

It will be impossible to extend our argument using the Mazur identity
to the six or higher dimensional Myers-Perry black hole solutions.
An $n$-dimensional space-time admitting $(n-3)$ commuting Killing
vector fields is always described by nonlinear $\sigma$-model
as shown by Maison~\cite{Maison:kx}.
To derive the Mazur identity for this nonlinear $\sigma$-model,
all the $(n-3)$ Killing vector fields have to be spacelike.
However, the $n$-dimensional Myers-Perry black hole space-time has only 
$[(n-1)/2]$ commuting spacelike Killing vector fields.
Thus our method cannot be used except for the five-dimensional
Myers-Perry black hole.

The rigidity theorem in four dimensions claims that the
asymptotically flat, stationary analytic space-time is also
axi-symmetric~\cite{Hawking:1971vc}.
However the existence of additional space-time Killing vector fields is not justified
in the case of
five-dimensional black holes any longer.
Therefore uniqueness shown in the present work
does not exclude the possibility of existence of
the black hole solutions with less symmetry as suggested by Reall~\cite{Reall:2002bh}.

\section*{Acknowledgments}
The authors would like to thank H.~Kodama for valuable discussions
and comments.
D.I. was supported by JSPS Research, and this research was supported in part by the
Grant-in-Aid for Scientific Research Fund (No. 6499).

\end{document}